\documentclass[referee]{raa}
\usepackage{graphicx,times}
\usepackage{natbib}
\usepackage{amssymb,amsmath}
\usepackage{mathrsfs}
\usepackage{color}
\usepackage{colordvi}
\usepackage{ulem}
\RequirePackage{lineno}

\usepackage[a4paper=true,dvipdfm=true,pagebackref=true]{hyperref}
\hypersetup{pdftitle = The title of my PDF, pdfauthor = My name, pdfsubject= The subject, pdfkeywords = keyword1 keyword2 keyword3}
\hypersetup{colorlinks = true, linkcolor = green, anchorcolor = red, citecolor = blue, filecolor = red, pagecolor = red, urlcolor = red}

\begin{document}

   \title{Smoothing methods comparison for CMB E- and B-mode separation
}

 \volnopage{ {\bf 2012} Vol.\ {\bf X} No. {\bf XX}, 000--000}
   \setcounter{page}{1}

   \author{Yi-Fan Wang\inst{}, Kai Wang*\email{ljwk@mail.ustc.edu.cn}\inst{}, Wen Zhao\inst{}
   }

   \institute{
   CAS Key Laboratory for Research in Galaxies and Cosmology,
   Department of Astronomy, University of Science
and Technology of China, Hefei, Anhui, 230026, China;\\
\vs \no
   {\small Received ...; accepted ...}
}

\abstract{ The anisotropies of the B-mode polarization in the cosmic microwave background radiation play a crucial role for the study of the very early Universe. However, in the real observation, the mixture of the E-mode and B-mode can be caused by the partial sky surveys, which must be separated before applied to the cosmological explanation. The separation method developed by Smith (\citealt{PhysRevD.74.083002}) has been widely adopted, where the edge of the top-hat mask should be smoothed to avoid the numerical errors. In this paper, we compare three different smoothing methods, and investigate the leakage residuals of the E-B mixture. We find that, if the less information loss is needed and the smaller region is smoothed in the analysis, the \textit{sin}- and \textit{cos}-smoothing methods are better. However, if we need a clean constructed B-mode map, the larger region around the mask edge should be smoothed. In this case, the \textit{Gaussian}-smoothing method becomes much better. In addition, we find that the leakage caused by the numerical errors in the \textit{Gaussian}-smoothing method mostly concentrates on two bands, which is quite easy to be reduced for the further E-B separations.
\keywords{Cosmic Microwave Background radiation, Polarization, Statistics
}
}

   \authorrunning{Yi-Fan Wang, Kai Wang, Wen Zhao}            
   \titlerunning{Smoothing methods comparison for CMB E- and B-mode separation}  
   \maketitle

%
\section{Introduction}           
\label{sect:intro}

\quad Cosmic microwave background(CMB) radiation encoded the fruitful information of the modern cosmology, which plays the crucial role for the determination of cosmological parameters (\citealt{aghanim2015planck}). The fluctuations of CMB include three parts: the temperature anisotropies, the E-mode polarization and the B-mode polarization. By the precise observations of WMAP and Planck satellites, the  prefect CMB temperature anisotropies, E-mode polarization and their correlation power spectra have been well observed (\citealt{aghanim2015planck}). However, the detection of B-mode is still quite awful (\citealt{ade2015joint}; \citealt{ade2014evidence}; \citealt{naess2014atacama}; \citealt{hanson2013detection}), which is the main goal of the next generation experiments (\citealt{bock604101task}). In the standard model, the CMB B-mode can only be generated by primordial gravitational waves and the cosmic weak lensing (second order effect). So, it provides the unique opportunity to directly probe the physics of the very early Universe through the primordial gravitational waves (\citealt{kamionkowski1997probe}; \citealt{seljak1997signature}).

  Comparing with the CMB temperature anisotropies and the E-mode polarization, the amplitude of the B-mode polarization is much smaller. Its detection is limited by various contaminations, including the foreground radiation, instrumental noises and instrumental systematics, and the E-B mixtures due to the partial sky surveys (\citealt{bock604101task}). In this paper, we shall focus on the E-B mixtures problem. Numerous numerical methods have been developed to separate the E- and B-mode polarization from the observable Q and U polarization maps. Among them, the methods developed by Smith \& Zaldarriaga (SZ method) (\citealt{PhysRevD.74.083002}; \citealt{PhysRevD.76.043001}), Zhao \& Baskaran (ZB method) (\citealt{PhysRevD.82.023001}), and Kim \& Naselsky (KN method) (\citealt{kim2010b}) have used the so-called $\chi$-field framework, and can be effectively applied to the potential data analysis or numerical simulations. In each method, in order to reduce the numerical errors, the usual top-hat CMB masks should be revised to the smoothed masks, where the edges of the masks are smoothly joint. In this work, we will compare three different smoothing methods adopted in different literatures, and investigate the residuals of the E-B mixtures  in those three methods. By this comparison, we will search for the best smoothing method, which induces the smallest leakage of the E-B separation.


\section{standard procedure for E- and B-mode separation and pure pseudo-$C_{\lowercase{l}}$ method}
\label{sect:standard procedure}

\quad First we present a brief review of two related definitions of E- and B-mode. Since CMB polarization does not contain the circular polarization component, it can be characterized completely by stokes parameters Q and U (\citealt{PhysRevD.55.7368}). Introduce the complex conjugate polarization fields $P_{\pm}$ defined as follows:
\begin{equation}
P_{\pm}\left(\hat{n}\right)\equiv Q\left(\hat{n}\right)\pm iU\left(\hat{n}\right),
\end{equation}
where $ \hat n $ denotes the direction of 2-dimensional sphere. It can be proved that the fields $ P_{\pm} $ have spin$ {\pm} $2, which means when rotate the coordinate system by an arbitrary angle $\alpha$ on the plane perpendicular to direction $ \hat{n} $, the polarization fields would change into:
\begin{equation}
P_{\pm}^{'}\left(\hat{n}\right)=P_{\pm}\left(\hat{n}\right)e^{\mp2i\alpha}.
\end{equation}

It is easier to study scalar field rather than spin-weighted field. One can construct such electric type and magnetic type scalar fields through Fourier expansion of $ P_{\pm} $ and a recombination of expansion coefficients. Expand $ P_{\pm} $ over spin-weighted spherical harmonic function bases (\citealt{PhysRevD.55.1830}):
\begin{equation}
P_{\pm}\left(\hat{n}\right)= \sum_{lm} a_{\pm 2,lm} \sideset{_{\pm2}}{_{lm}}{\mathop{\mathrm{Y}}}\left(\hat n\right),
\end{equation}
where $ \sideset{_{\pm2}}{_{lm}}{\mathop{\mathrm{Y}}} $ are $ \pm 2 $ spin-weighted spherical harmonic functions. And one can calculate multipole coefficients $ a_{\pm 2,lm} $ as:
\begin{equation}
a_{\pm 2,lm}=\int d\hat n P_{\pm}\left(\hat{n}\right) \sideset{_{\pm2}}{_{lm}^{\ast}}{\mathop{\mathrm{Y}}}\left(\hat n\right).
\end{equation}
The coefficients of scalar E and B fields are defined as a recombination of $ a_{\pm 2,lm} $:
\begin{align}
E_{lm}&\equiv-\frac{1}{2}[a_{2,lm}+a_{-2,lm}],\nonumber\\
B_{lm}&\equiv-\frac{1}{2i}[a_{2,lm}-a_{-2,lm}].
\end{align}
Then the E- and B-mode of polarization field are defined as,
\begin{equation}
E(\hat n)\equiv \sum_{lm}E_{lm}Y_{lm}(\hat n), \qquad
B(\hat n)\equiv \sum_{lm}B_{lm}Y_{lm}(\hat n),
\end{equation}
and the power spectra are defined as,
\begin{align}
C_{l}^{EE}\equiv\frac{1}{2l+1}\sum_{m}\left<E_{lm}E_{lm}^{\ast}\right>,\nonumber\\
C_{l}^{BB}\equiv\frac{1}{2l+1}\sum_{m}\left<B_{lm}B_{lm}^{\ast}\right>.
\end{align}
where the brackets denote the ensemble average. Since the E and B fields are Gaussian random fields, the power spectra defined above encode all the statistical information of the fields.

Another related definition is to use spin lowering and raising operators to construct electric type and magnetic type scalar fields (\citealt{PhysRevD.55.1830})
\begin{equation}
\mathcal{E}(\hat n)\equiv -\frac{1}{2}[\bar {\eth} \bar{\eth}P_{+}(\hat n)+{\eth}{\eth}P_{-}(\hat n)],
\end{equation}
\begin{equation}\label{bfield}
\mathcal{B}(\hat n)\equiv -\frac{1}{2i}[\bar {\eth} \bar{\eth}P_{+}(\hat n)-{\eth}{\eth}P_{-}(\hat n)].
\end{equation}
$ \bar {\eth} $ and $ {\eth} $ are the spin lowering and raising operators respectively, which are defined as follows:
\begin{equation}
\bar\eth f\equiv-\sin^{-s}{\theta} \bigg(\frac{\partial}{\partial\theta}-\frac{i}{\sin\theta}\frac{\partial}{\partial\phi}\bigg)(f\sin^{s}\theta),
\end{equation}
\begin{equation}
\eth f\equiv-\sin^{s}{\theta} \bigg(\frac{\partial}{\partial\theta}+\frac{i}{\sin\theta}\frac{\partial}{\partial\phi}\bigg)(f\sin^{-s}\theta),
\end{equation}
where $f$ is an arbitrary function with spin $s$.
Decompose the $ \mathcal{E} $ and $ \mathcal{B} $ fields over the scalar spherical harmonics bases:
\begin{equation}
\mathcal E(\hat n)\equiv \sum_{lm}\mathcal E_{lm}Y_{lm}(\hat n), \qquad
\mathcal B(\hat n)\equiv \sum_{lm}\mathcal B_{lm}Y_{lm}(\hat n),
\end{equation}
where the decomposition coefficients are calculated as:
\begin{equation}
\mathcal{E}_{lm}=\int d\hat{n} \mathcal{E}(\hat{n})Y_{lm}^{\ast}(\hat{n}), \qquad
\mathcal{B}_{lm}=\int d\hat{n} \mathcal{B}(\hat{n})Y_{lm}^{\ast}(\hat{n}).
\end{equation}
Construct the power spectra as the same manner of the former method:
\begin{align}
C_{l}^{\mathcal{E}\mathcal{E}}\equiv\frac{1}{2l+1}\sum_{m}\left<\mathcal{E}_{lm}\mathcal{E}_{lm}^{\ast}\right>,\nonumber\\
C_{l}^{\mathcal{B}\mathcal{B}}\equiv\frac{1}{2l+1}\sum_{m}\left<\mathcal{B}_{lm}\mathcal{B}_{lm}^{\ast}\right>.
\end{align}
Thanks to a property of spin lowering and raising operators:
\begin{align}\label{spin-harmonic relation}
\bar{\eth}\sideset{_s}{_{lm}}{\mathop{\mathrm{Y}}}(\hat{n})=-\sqrt{(l+s)(l-s+1)}\sideset{_{s-1}}{_{lm}}{\mathop{\mathrm{Y}}}(\hat{n}),\nonumber\\
{\eth}\sideset{_s}{_{lm}}{\mathop{\mathrm{Y}}}(\hat{n})=\sqrt{(l-s)(l+s+1)}\sideset{_{s+1}}{_{lm}}{\mathop{\mathrm{Y}}}(\hat{n}),
\end{align}
the relationships of the multipoles and power spectra between these two definitions are:
\begin{equation}
\mathcal{E}_{lm}=N_{l,2}E_{lm},\qquad
\mathcal{B}_{lm}=N_{l,2}B_{lm},
\end{equation}
\begin{equation}
C_l^{\mathcal{E}\mathcal{E}}=N_{l,2}^{2}C_{l}^{EE},\qquad
C_l^{\mathcal{B}\mathcal{B}}=N_{l,2}^{2}C_{l}^{BB},
\end{equation}
where $N_{l,s}=\sqrt{(l+s)!/(l-s)!} $ .

In the actual observation, we must mask out a fractional portion of sky due to the foreground contamination. The observed values of Stokes parameters $ \tilde{Q} $ and $ \tilde{U} $ are related to the real values $Q$ and $U$ by introducing a window function $W(\hat{n})$:
\begin{equation}
\tilde{Q}=QW,\qquad
\tilde{U}=UW.
\end{equation}
The value of $W$ is non-zero only in the observational region of sky.

However, if one apply $ \tilde{P}_{\pm}=\tilde{Q}\pm  i\tilde{U} $ directly to the above two definitions of E- and B-mode, it will lead to the so-called E-B mixture problem arising from cut-sky effect (\citealt{PhysRevD.65.023505}; \citealt{PhysRevD.67.023501}), dramatically restricting the detectability of B-mode signal. Several methods were brought up to solve this problem (\citealt{PhysRevD.68.083509}; \citealt{PhysRevD.65.023505}; \citealt{PhysRevD.67.023501}; \citealt{bunn2008b}; \citealt{PhysRevD.79.123515}; \citealt{PhysRevD.74.083002}; \citealt{PhysRevD.76.043001}; \citealt{PhysRevD.78.123533}; \citealt{cao2009wavelet}; \citealt{PhysRevD.82.023001};\citealt{kim2010b}). We notice that the article (\citealt{PhysRevD.88.023524}) have compared three different methods of them which are numerically fast enough (\citealt{PhysRevD.74.083002}; \citealt{PhysRevD.76.043001}; \citealt{PhysRevD.82.023001}; \citealt{kim2010b}) and drew the conclusion that SZ method (\citealt{PhysRevD.74.083002};  \citealt{PhysRevD.76.043001}) is the best in the meaning of significantly reducing E to B leakage and ensuring the smallest error bars at the same time. Therefore we choose to apply this best method to the E- and B-mode separating operation in the following paper.

We briefly review how the SZ method separate E- and B-mode on an incomplete sky. First the concept of pure pseudo-multipoles is put forward and defined as:
\begin{align}\label{pure pseudo-multipoles}
\mathcal{E}_{lm}^{pure}\equiv-\frac{1}{2}\int d\hat{n}\left\{P_{+}(\hat n)\left[\bar {\eth} \bar{\eth}\left(W(\hat{n})Y_{lm}(\hat{n})\right)\right]^\ast+P_{-}(\hat n)\left[ {\eth} {\eth}\left(W(\hat{n})Y_{lm}(\hat{n})\right)\right]^\ast \right\},\nonumber\\
\mathcal{B}_{lm}^{pure}\equiv-\frac{1}{2i}\int d\hat{n}\left\{P_{+}(\hat n)\left[\bar {\eth} \bar{\eth}\left(W(\hat{n})Y_{lm}(\hat{n})\right)\right]^\ast-P_{-}(\hat n)\left[ {\eth} {\eth}\left(W(\hat{n})Y_{lm}(\hat{n})\right)\right]^\ast \right\}.
\end{align}
Recall the definition of pseudo-multipoles which are concentrated in the pseudo-$ C_{l} $ estimator technique (\citealt{efstathiou2004myths}):
\begin{equation}\label{pseudo-multipoles}
\tilde{\mathcal{E}}_{lm}\equiv\int d\hat{n} \mathcal{E}(\hat{n})W(\hat{n})Y_{lm}^{\ast}(\hat{n}), \qquad
\tilde{\mathcal{B}}_{lm}\equiv\int d\hat{n} \mathcal{B}(\hat{n})W(\hat{n})Y_{lm}^{\ast}(\hat{n}).
\end{equation}
It can be proved (\citealt{PhysRevD.76.043001}) the expressions of Eq.(\ref{pure pseudo-multipoles}) and Eq.(\ref{pseudo-multipoles}) are equivalent. This shows that, in principle the pure pseudo-multipoles method can successfully extract the pure E- and B-mode signal and avoid E-B mixing part. To calculate the expression of Eq.(\ref{pure pseudo-multipoles}), one needs to use the property of spin raising and lowering operators:
\begin{equation}\label{prop1}
\bar \eth \left(fg\right )=\left (\bar \eth f\right )g+f\left (\bar \eth g\right ), \qquad
\eth \left(fg\right )=\left ( \eth f\right )g+f\left ( \eth g\right ),
\end{equation}
where $ f $ and $ g $ are arbitrary spin-weighted functions with spin $ s_1 $ and $ s_2 $, and $ f\!g $ is spin $ s_1\!+\!s_2 $ weighted function, together with the complex conjugate relationship $\eth^{\ast}=\bar{\eth}$.
Use the Eqs.(\ref{spin-harmonic relation}, \ref{prop1}), and substitute them into Eq.(\ref{pure pseudo-multipoles}). one finally obtains (only focus on B-mode):
\begin{align}\label{pureb}
\mathcal{B}_{lm}^{pure}=-\frac{1}{2i}\int d \hat{n} \bigg{[}
& P_{+}\bigg{(} \left(\bar {\eth} \bar{\eth}W\right) Y_{lm}^{\ast}+2N_{l,1} \left(\bar{\eth } W\right)  \left(\sideset{_{1}}{_{lm}^{\ast}}{\mathop{\mathrm{Y}}}\right)+N_{l,2} W \left(\sideset{_{2}}{_{lm}^{\ast}}{\mathop{\mathrm{Y}}}\right)\bigg{)} \\
 -& P_{-}\bigg ( \left({\eth}{\eth}W \right)Y_{lm}^{\ast}-2N_{l,1}\left({\eth } W\right ) \left( \sideset{_{-1}}{_{lm}^{\ast}}{\mathop{\mathrm{Y}}}\right)+N_{l,2} W\left ( \sideset{_{-2}}{_{lm}^{\ast}}{\mathop{\mathrm{Y}}}\right)\bigg )\nonumber
\bigg{]},
\end{align}
where:
\begin{equation}
\eth W=-\frac{\partial W}{\partial \theta }-\frac{i}{\sin \theta }\frac{\partial W}{\partial \phi},
\end{equation}
\begin{equation}
\eth \eth W=-\cot \theta \frac{\partial W}{\partial \theta }+\frac{\partial^2 W}{\partial \theta^2 }-\frac{1}{\sin^2 \theta}\frac{\partial ^2 W}{\partial \phi^2}-\frac{2i\cot \theta}{\sin \theta }\frac{\partial W}{\partial \phi}+\frac{2i}{\sin \theta}\frac{\partial^2 W}{\partial \theta \partial \phi}.
\end{equation}
The Eq. (\ref{pureb}) is the basis for all following E- and B-mode separating operations.

\section{smoothing methods comparison}
\label{sect:smoothing methods comparison}
\begin{figure}
   \centering
   \includegraphics[width=14cm]{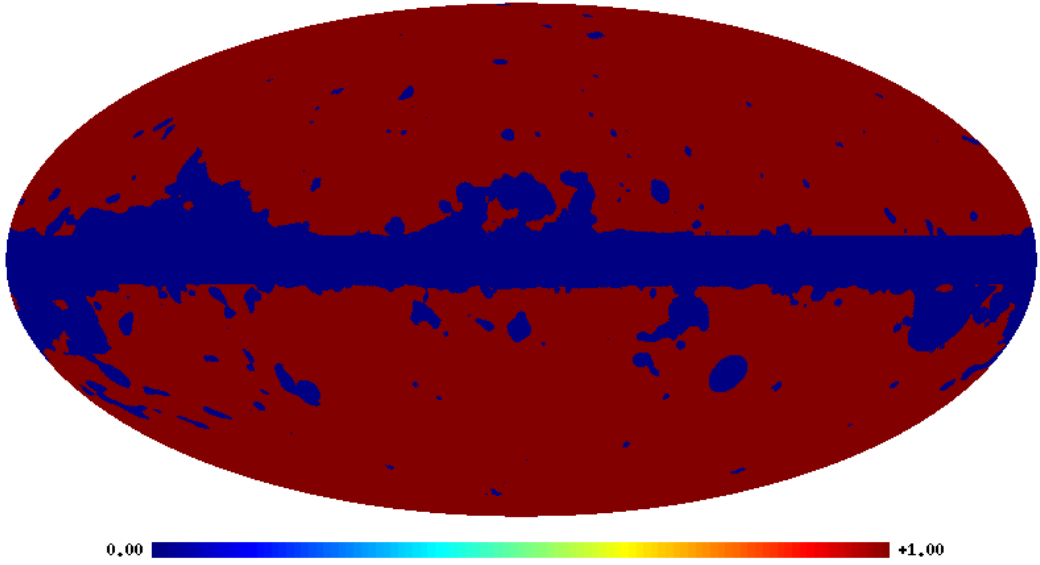}
   \caption{A window function for CMB polarization published by Planck collaboration. }
   \label{Fig1}
   \end{figure}
\quad In this section we shall discuss the effect on E- and B-mode separation brought by different choice of window function $ W(\hat{n}) $. The simplest case is $ W(\hat{n})=1 $ in the observational region of sky and $ W(\hat{n})=0 $ outside (referred to as ``top-hat window function'') as shown in Fig.\ref{Fig1} published by Planck collaboration. However, in the Eq. (\ref{pureb}), the calculation of derivative of $ W(\hat{n}) $ is inevitable, so we must smooth the edge of $ W(\hat{n}) $ (also called "apodization"). The noticed point is that the zero-value pixels of the original window function should be zero after being smoothed. Another restriction on the smoothing method of $ W(\hat{n}) $ given by SZ method (\citealt{PhysRevD.74.083002}) that both $ W $ and its gradient should vanish at the boundary of observational sky. We compare three different smoothing methods for the window functions, which have appeared in the literatures and satisfy the above conditions. The first two methods use trigonometric function to smooth the edge of top-hat window function, referred to as \textit{cos}-smoothing and \textit{sin}-smoothing (\citealt{PhysRevD.79.123515})  respectively. Their expressions are:
\begin{equation}\label{cos}
W_i=\left\{
\begin{aligned}
&{\frac{1}{2}-\frac{1}{2}\cos(\frac{\delta_i}{\delta_c}\pi)}  &\delta_{i}<\delta_{c}\\
&1  &\delta_{i}>\delta_{c}\\
\end{aligned}
\right.
\end{equation}
and
\begin{equation}\label{sin}
W_i=\left \{
\begin{aligned}
&-\frac{1}{2\pi}\sin\left( 2\pi \dfrac{\delta_{i}}{\delta_{c}}\right)+\frac{\delta_i}{\delta_c}  &\delta_{i}<\delta_{c}\\
&1  &\delta_{i}>\delta_{c}\\
\end{aligned}
\right.
\end{equation}
where $ \delta_i $ is the distance from each $1$-valued pixel to the closest 0-valued pixel in the top-hat window function, and $ \delta_c $ is a constant set in advance representing the smoothing range.

\begin{figure}
   \centering
   \includegraphics[width=12cm]{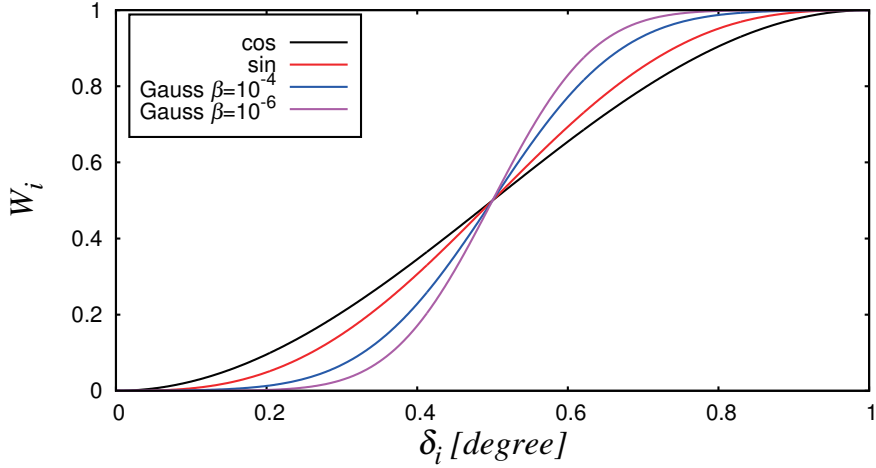}
   \caption{ The plot of \textit{cos}-smoothing, \textit{sin}-smoothing and \textit{Gaussian}-smoothing in real space.}
   \label{Fig2}
   \end{figure}

The third smoothing method is put forward by (\citealt{kim2011make}). This article analyzed the generation of  numerical error in E- and B-mode separation theoretically by introducing Gibbs phenomenon, which says the partial sum of the Fourier series of a function with jump discontinuities has large oscillations near the jump. Therefore one can reduce the Gibbs phenomenon by choosing a window function whose multipoles higher than the truncation point are suppressed. The author use Gaussian smoothing kernel to smooth the edge of window function whose expression is:
\begin{equation}\label{gauss}
W_i=\left \{
\begin{aligned}
&\int_{-\infty }^{\delta_i-\frac{\delta_c}{2}} \frac{1}{\sqrt{2\pi \sigma^2}}\exp\bigg(-\frac{x^2}{2\sigma^2}\bigg)dx=\frac{1}{2}+\frac{1}{2}{\rm erf}\bigg(\frac{\delta_i-\frac{\delta_c}{2}}{\sqrt{2}\sigma }\bigg)  &\delta_{i}<\delta_{c}\\
&1  &\delta_{i}>\delta_{c}\\
\end{aligned}
\right.
\end{equation}
where $ \sigma=\frac{\rm FWHM}{\sqrt{8\ln2}} $ and FWHM denotes the full width at half maximum of the smoothing kernel. Let $ \beta $ denotes the jump range at $\delta_i=\delta_c $ and $\delta_i=0 $, then:
\begin{equation}
\beta=\frac{1}{2}-\frac{1}{2}{\rm erf}\bigg(\frac{\frac{\delta_c}{2}}{\sqrt{2}\sigma }\bigg).
\end{equation}
The $\beta $ is a small and adjustable parameter. Set $ \delta_c=1^\circ$ and $\beta=10^{-4},10^{-6}$ respectively, plot the values of Eqs. (\ref{cos}, \ref{sin}, \ref{gauss}) in the Fig.\ref{Fig2}.

\begin{figure}
   \centering
   \includegraphics[width=14cm]{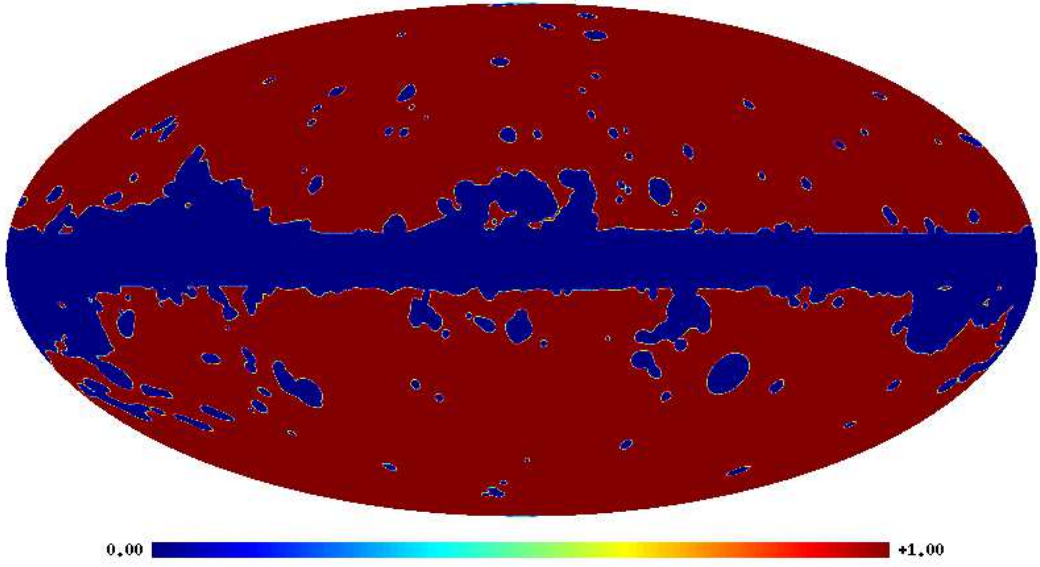}
   \caption{A smoothed window function using \textit{Gauss}-smoothing method with parameters $ \delta_c=1^{\circ} $ and $ \beta=10^{-4} $. }
   \label{smoothed}
   \end{figure}

\begin{figure}
    \centering
   \includegraphics[width=15cm,height=12cm]{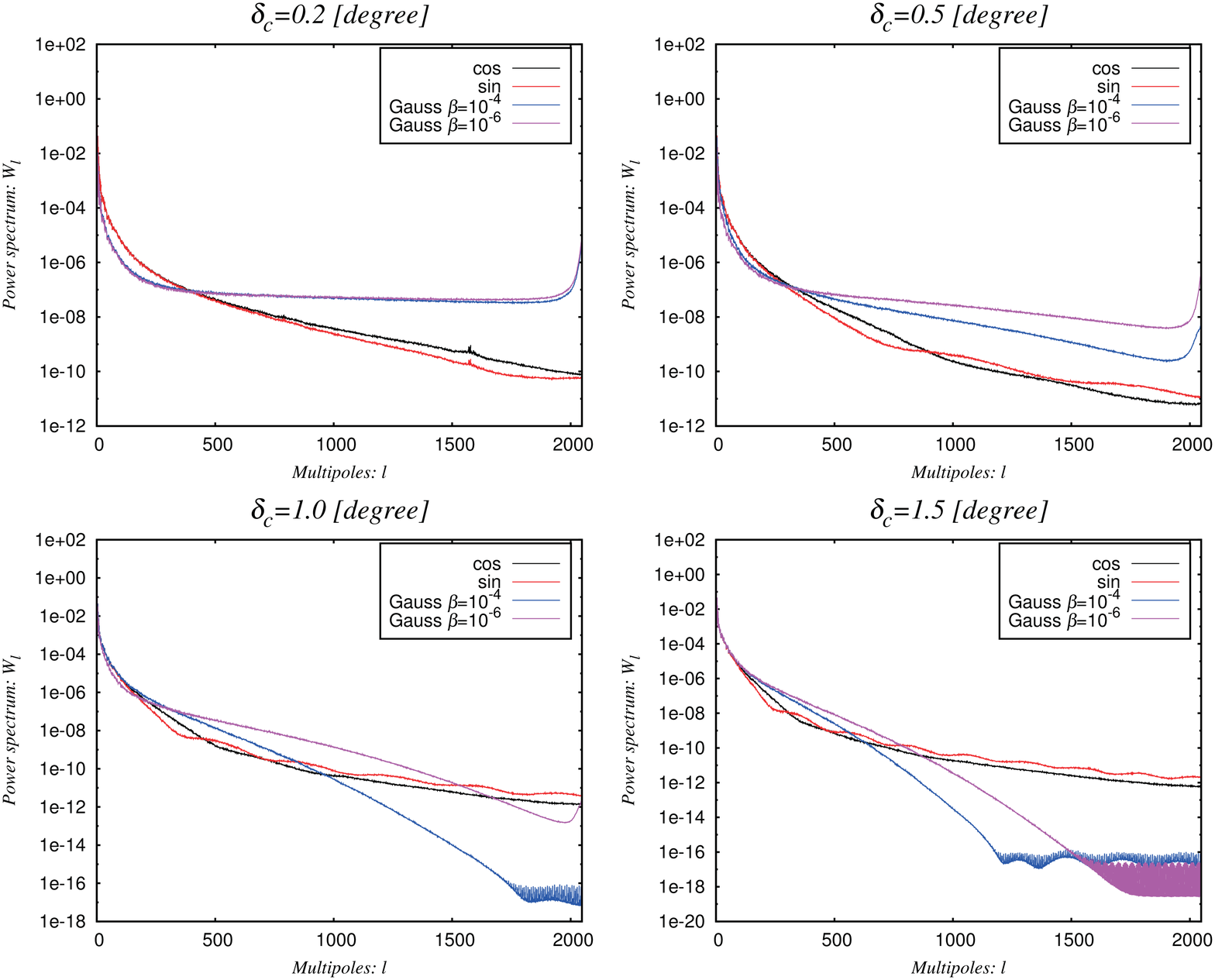}
   \caption{The power spectrum of smoothing window function $ W(\hat{n}) $ with different smoothing methods and different parameters.}
   \label{Fig3}
   \end{figure}

Inspired by the explanation of Gibbs phenomenon, we also analyze the smoothed window functions in harmonic space. Apply the \textit{cos}-smoothing, \textit{sin}-smoothing and \textit{Gaussian}-smoothing on the top-hat window function, then decompose the smoothed window function $ W(\hat{n}) $ (shown in Fig.\ref{smoothed}) on spherical harmonic basis and define the power spectrum  $ W_l $ as follows:
\begin{equation}
W_l=\frac{1}{2l+1}\sum_{m}w_{lm}w_{lm}^{\ast},
\end{equation}
where
\begin{equation}
w_{lm}=\int d \hat{n} W(\hat{n})Y_{lm}^{\ast}(\hat{n}).
\end{equation}
The power spectrum of smoothed window function using \textit{cos}, \textit{sin} and \textit{Gaussian} methods with different parameters is shown as Fig.\ref{Fig3}. We can see from the figure that the $ Gaussian $-smoothed window function has the lower power spectrum in small scale.

\begin{figure}
   \centering
   \includegraphics[width=7cm]{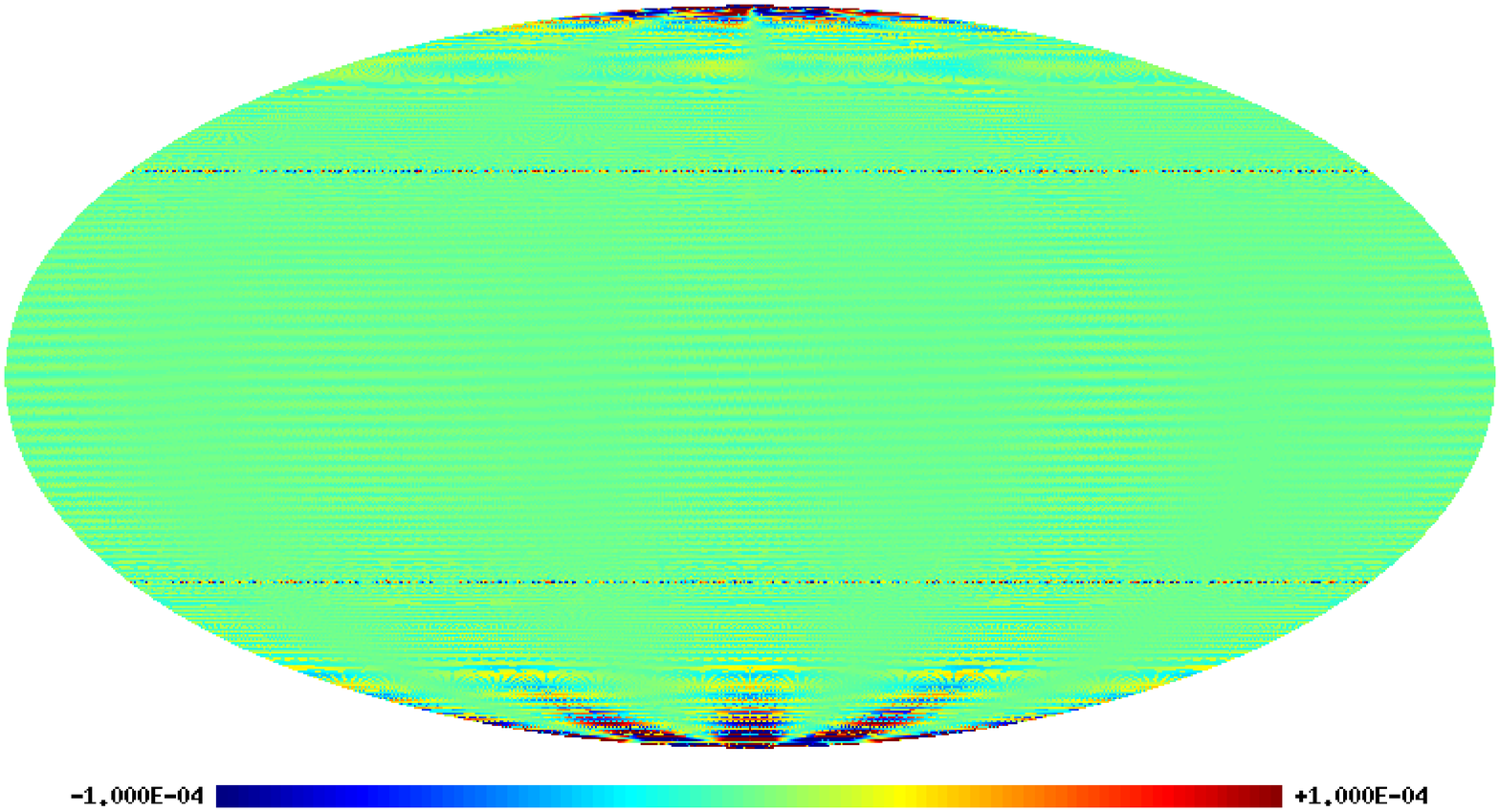}
         \includegraphics[width=7cm]{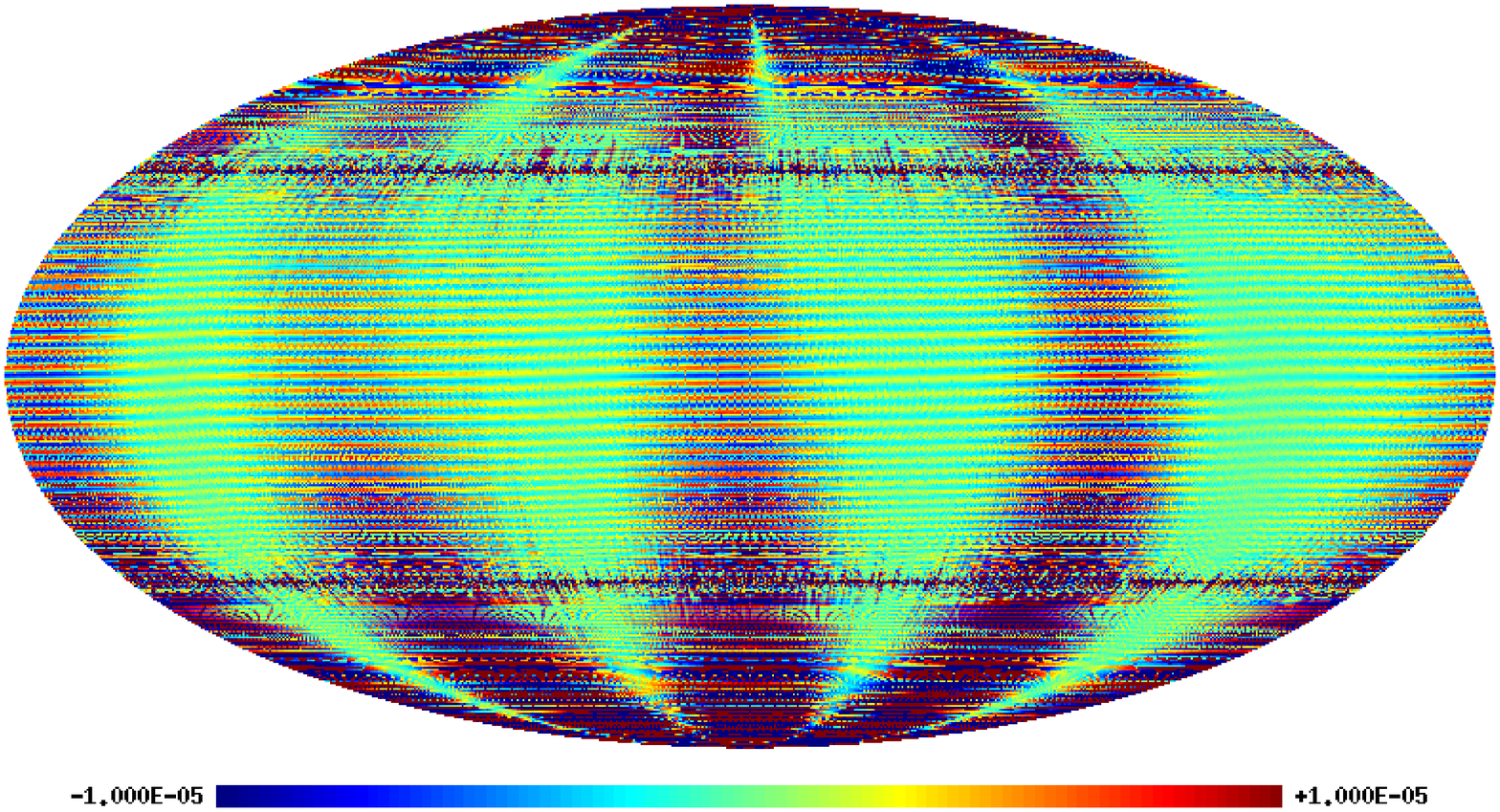}
   \includegraphics[width=7cm]{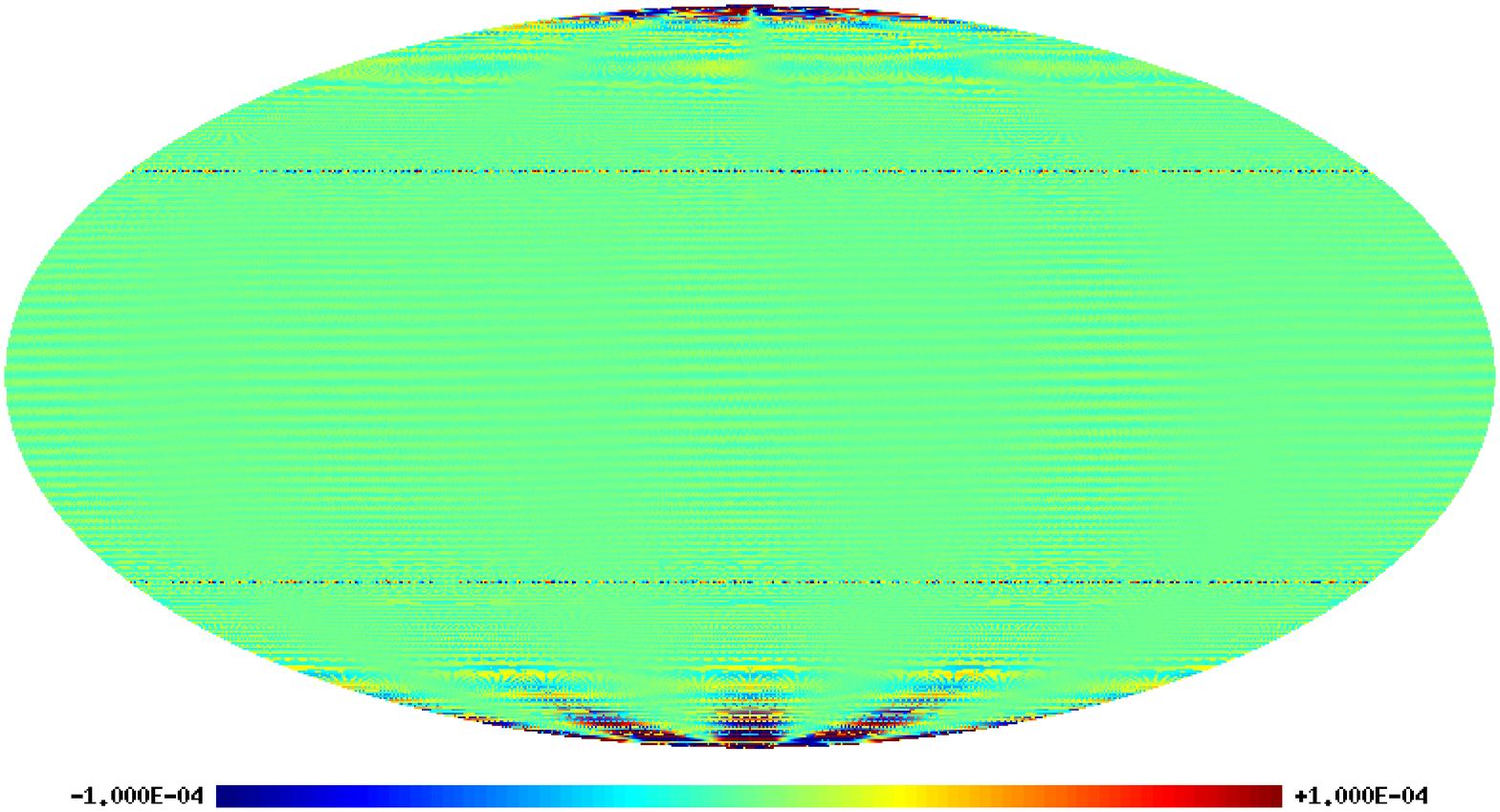}
      \includegraphics[width=7cm]{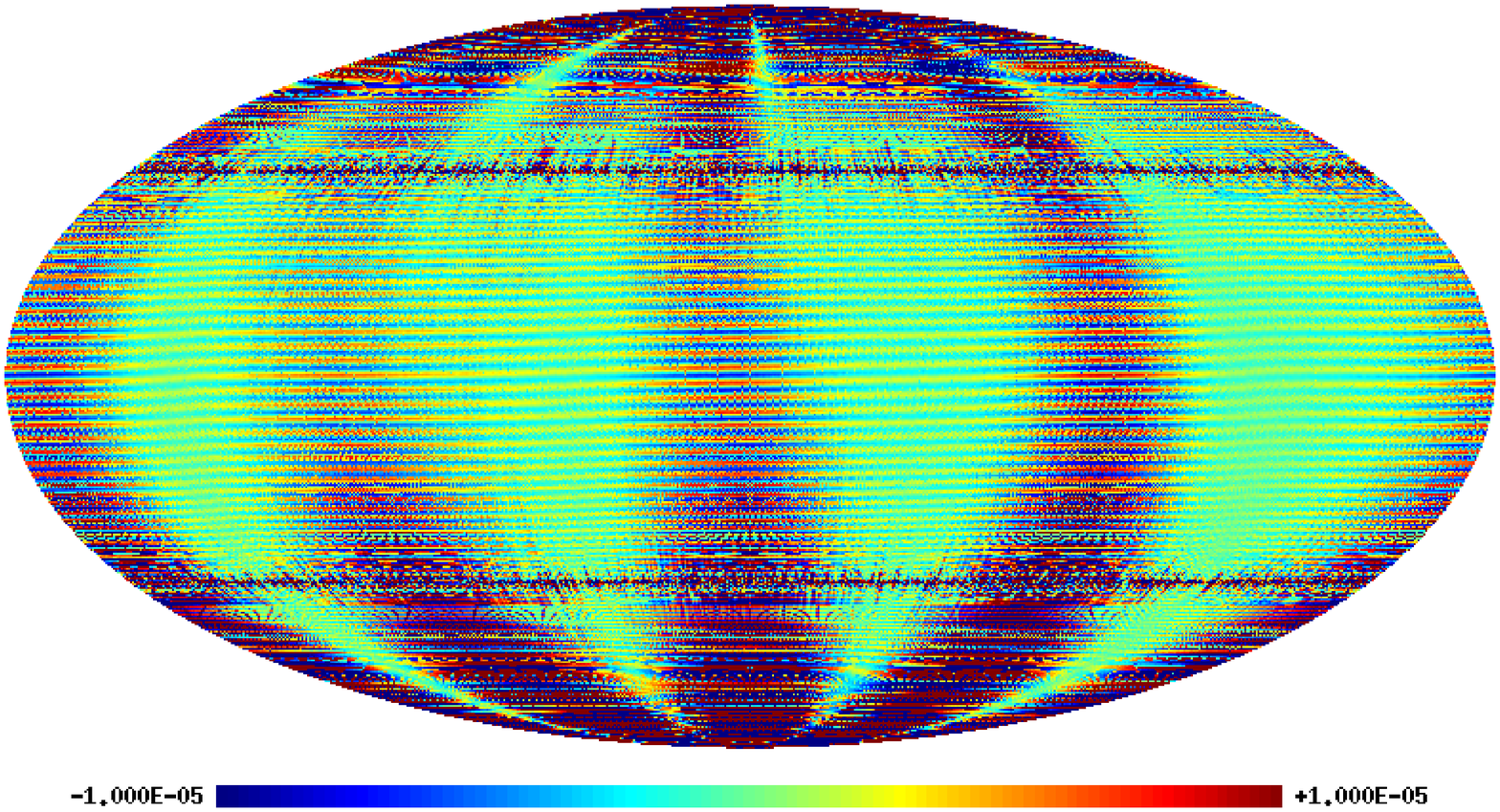}
      \includegraphics[width=7cm]{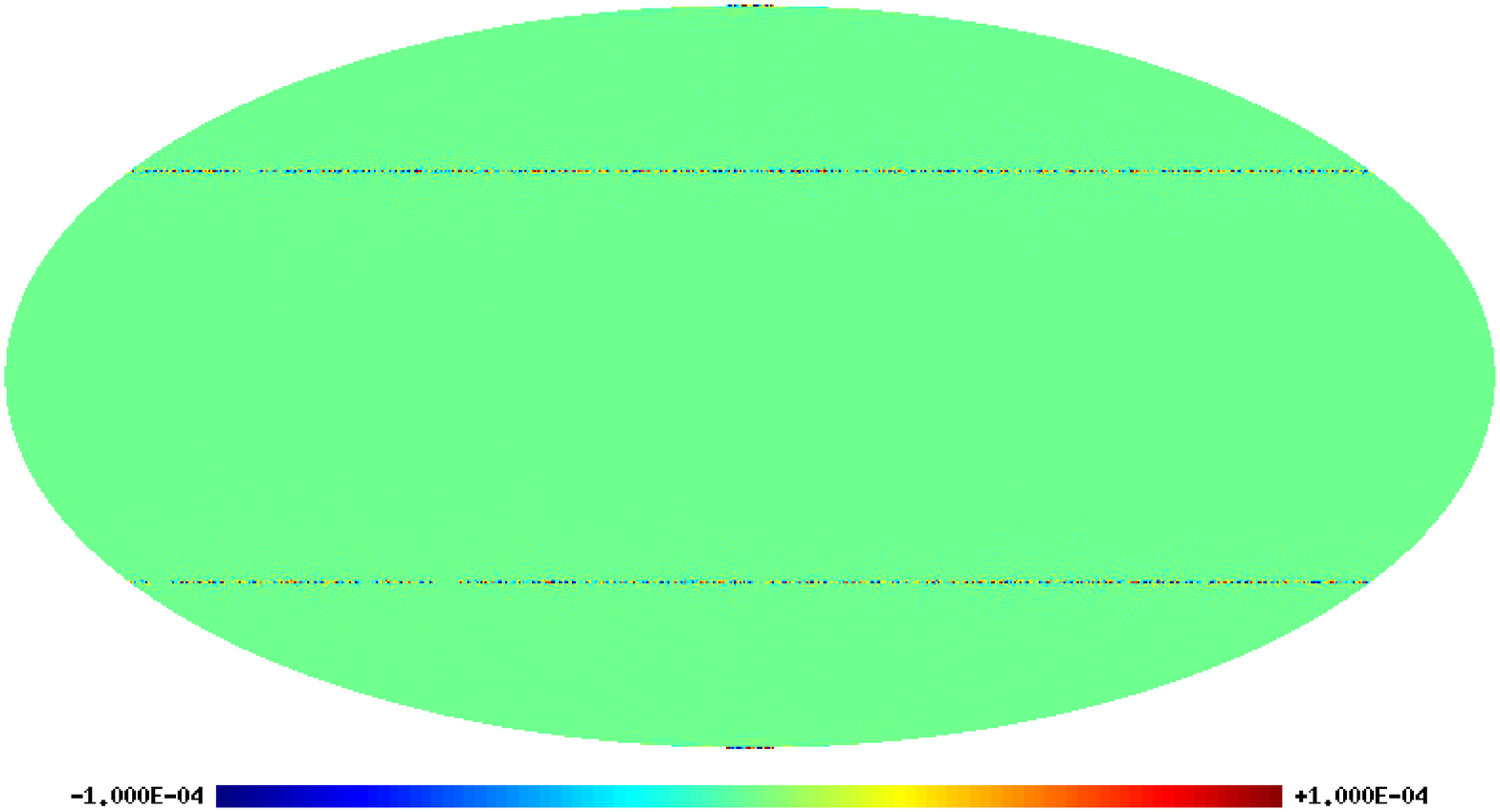}
   \includegraphics[width=7cm]{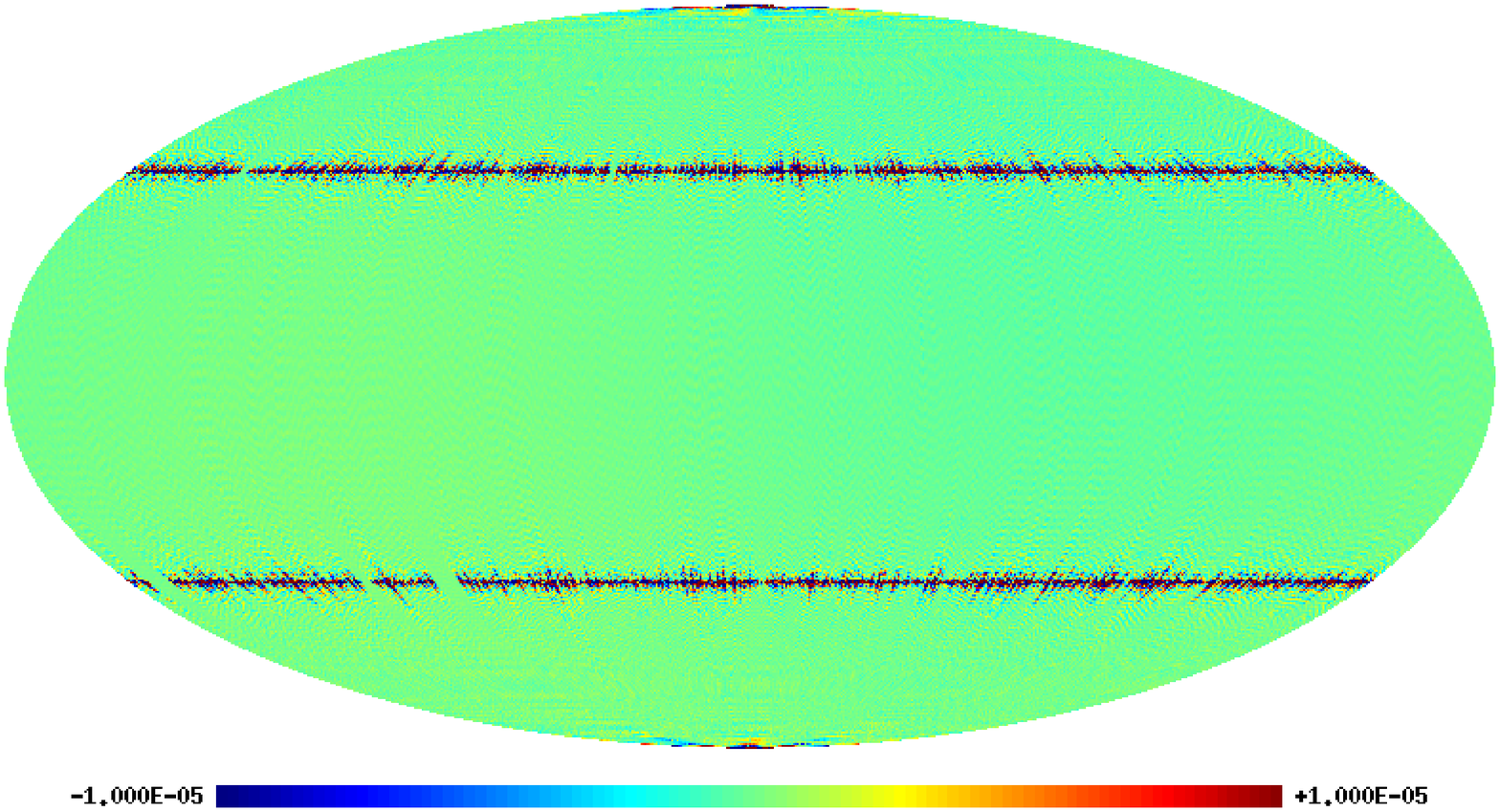}
   \caption{The pure B-mode field constructed by SZ method and \textit{cos}-, \textit{sin}- and \textit{Gaussian}-smoothing window function, from top to bottom, respectively, where $ \delta_c=1^{\circ} $ and $ \beta=10^{-4} $. The panels on the right side have the scaling magnified in order to show the residual leakage.}
   \label{figurebfield}
   \end{figure}

   We shall investigate the numerical error due to finite pixelization in E- and B-mode separation through simulated polarization maps. First use the \textit{synfast} subroutine in the HEALPix package to generate the full sky $ Q(\hat{n}) $ and $ U(\hat{n}) $ maps with the best-fit cosmological parameters published by Planck 2013:
   \begin{align}\label{parameter}
\Omega_{b}h^{2}&=0.022068,\Omega_{c}h^{2}=0.12029, \nonumber\\
\Omega_{\Lambda}&=0.6825,\tau_{reion}=0.0925, \nonumber\\
A_{s}&=2.215*10^{-9},n_s=0.9624.
\end{align}
Set the resolution of simulated maps with $ N_{side}=1024 $ and the Gaussian beam with $\theta_{F}=30'$. Assume no contribution from gravitational waves nor cosmic lensing, i.e. $ C_l^{BB}=0 $ in the input model. Then use SZ method and a specific smoothed window function to construct the pure B-mode field as:
\begin{equation}\label{purebfield}
\mathcal{B}^{pure}(\hat{n})=\sum_{lm}\mathcal{B}^{pure}_{lm}Y_{lm}(\hat{n}),
\end{equation}
where the expression of $ \mathcal{B}^{pure}_{lm} $ is shown in Eq.(\ref{pureb}). The $ \mathcal{B}^{pure}(\hat{n}) $ is related to $ \mathcal{B}(\hat{n}) $ in Eq.(\ref{bfield}) by
\begin{equation}
 \mathcal{B}^{pure}(\hat{n})=\mathcal{B}(\hat{n})W(\hat{n}).
\end{equation}
The Fig.\ref{figurebfield} is a visualization of $ \mathcal{B}^{pure}(\hat{n}) $. Since we assume $ C_l^{BB}=0 $, all the non-zero value pixels in Fig.\ref{figurebfield} are attributed to the numerical error.

\begin{figure}
   \centering
   \includegraphics[width=7cm]{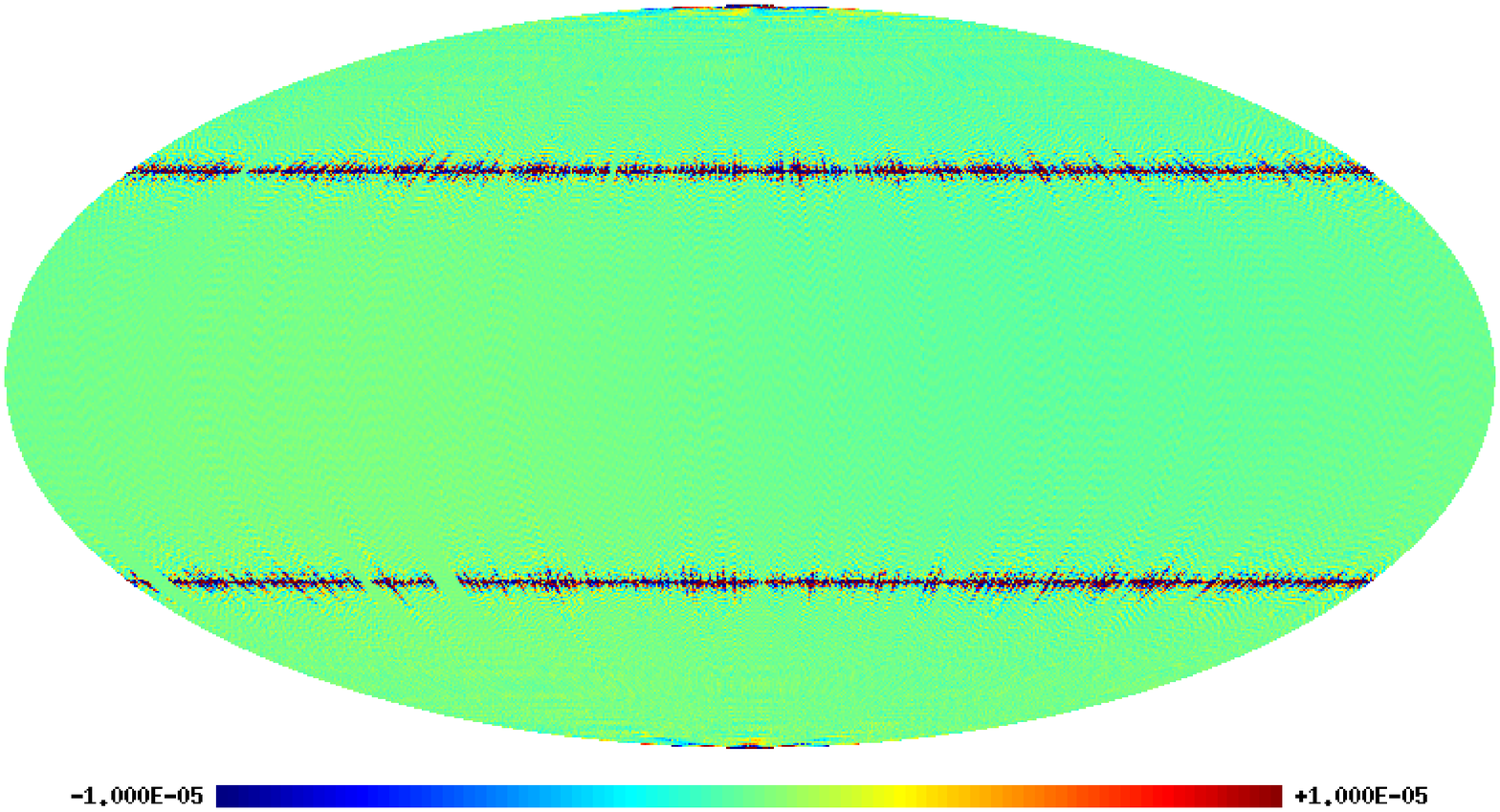}
   \includegraphics[width=7cm]{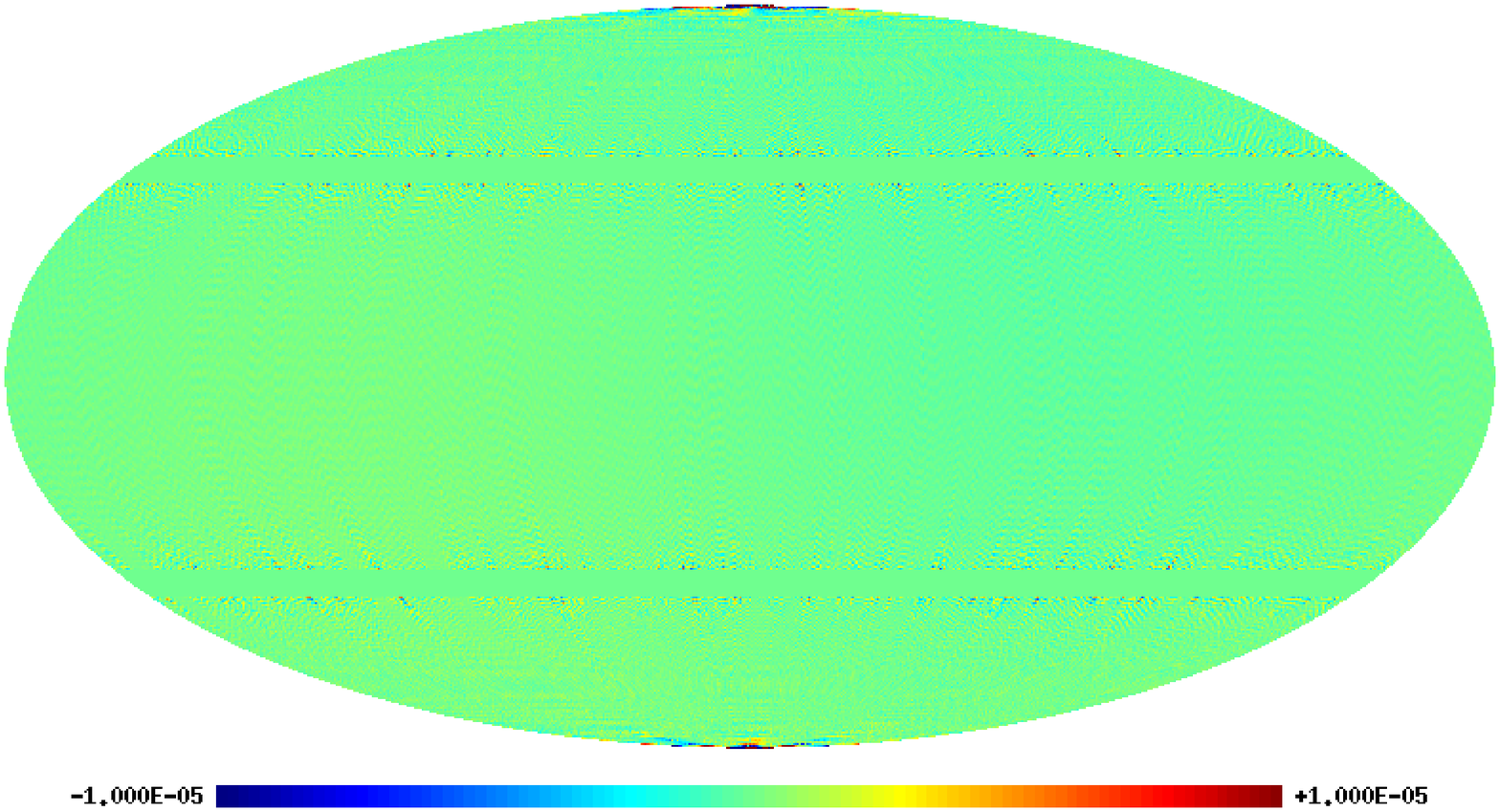}
   \caption{The pure B-mode field constructed by SZ method with \textit{Gaussian}-smoothed window function. The right one has been masked out the contamination bands, and the left one has not, with the same plotting scale.}
   \label{bandmask}
   \end{figure}

Interesting enough, the third panel in Fig.\ref{figurebfield} shows that the numerical errors mostly concentrate on two bands, due to the program design of HEALPix package. The HEALPix package divides the sky into three parts, and resembles them after operation, so there will be some residue on the joint. Besides, due to this kind of residue locates in two narrow bands, we can mask them out to remove most of the contamination with little information lost. In Fig.\ref{bandmask}, We mask out two bands centered at $48^{\circ}$  and $132^{\circ}$  and the width of each band is $6^{\circ}$, then the map looks much cleaner. How to quantify this further reduction on the numerical errors in the constructed pure B-mode map is another important topic in this area, we leave it as a new work.

\begin{figure}
   \centering
   \includegraphics[width=15cm,height=12cm]{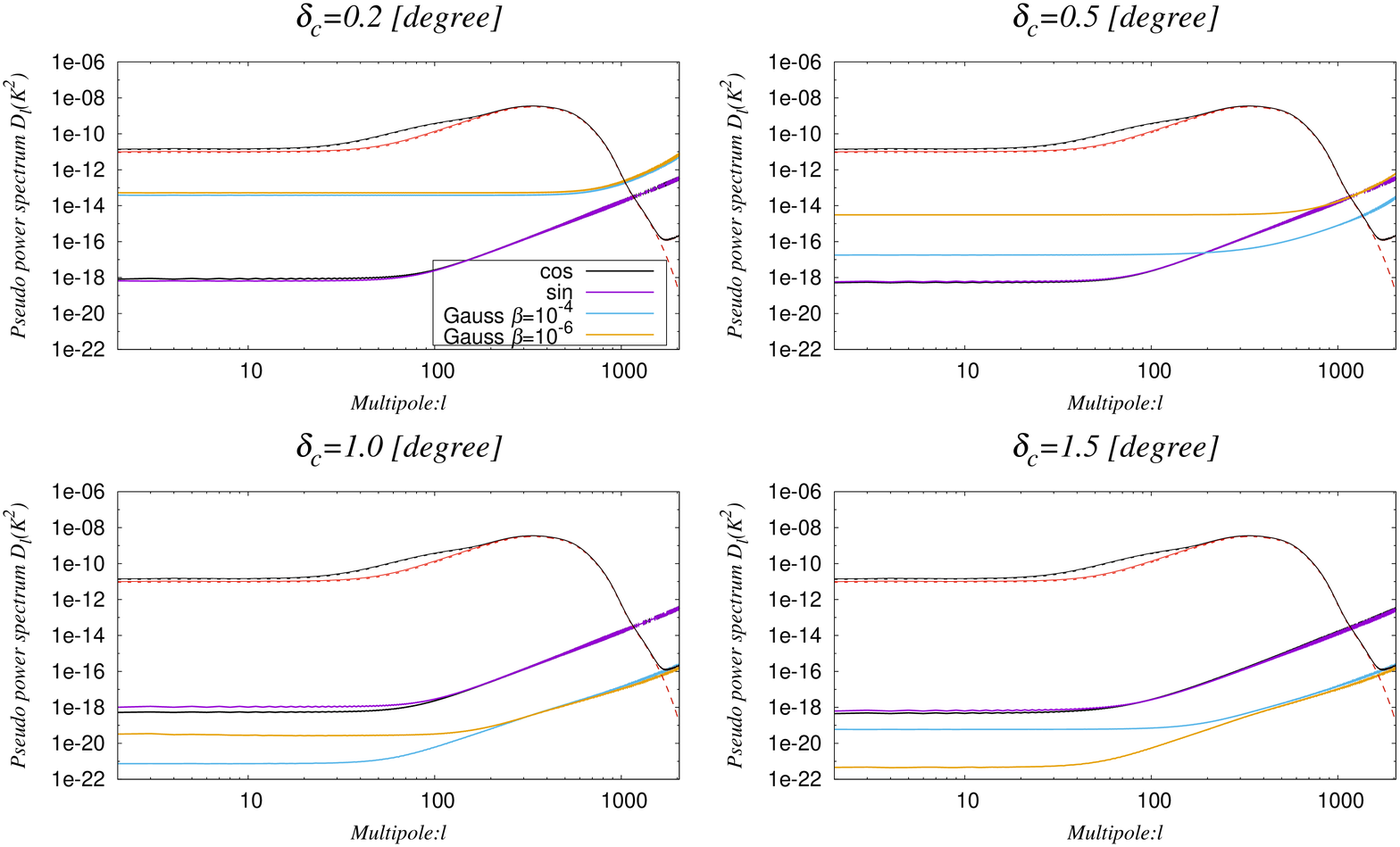}
   \caption{The pseudo power spectrum of $ \mathcal{B}^{pure}(\hat{n}) $ obtained using different smoothed window functions. We also plot the power spectrum with signal, using the smoothed mask with $\delta_c=1.0^{\circ}$ and $\beta=10^{-4}$, which are the theoretical power spectrum with $r=0.1$ (black dash line) and $r=0$ (red dash line), and the average power spectrum of 1000 simulations with $r=0.1$ (black solid line) and $r=0$ (red solid line). All of the four extra lines are with contribution of cosmic lensing.}
   \label{mainresult}
   \end{figure}

In order to quantify the numerical errors of pure B-mode map in harmonic space, we define the pseudo power spectrum as:
\begin{equation}
D_l^{pure}=\frac{1}{2l+1}\sum_{m}\mathcal{B}_{lm}^{pure}\mathcal{B}_{lm}^{pure\ast}.
\end{equation}
We use Monte Carlo method to investigate the effect on numerical errors using different smoothed window functions. The result is shown in Fig.\ref{mainresult}. Each line is an average over 500 realizations with the same cosmological initial conditions but different random seed. To compare the magnitude of signal with numerical error, we also plot the theoretical pseudo power spectrum (see Eq.(38) in \citealt{PhysRevD.82.023001}) and the average power spectrum of 1000 simulations with initial condition $ r=0 $ and $ r=0.1 $ ($r$ is primordial tensor-to-scalar ratio) respectively and all with the contribution of cosmic lensing.

   In the Fig.\ref{mainresult}, for the  $ \mathcal{B}^{pure}(\hat{n}) $ with initial conditions $ C_l^{BB}=0 $, all the non-zero values of pseudo power spectra are due to numerical error. Therefore, it can be used to measure the intensity of contamination quantitatively. We can recognize that the tendency of the pseudo power spectrum brought by different smoothed window functions in Fig.\ref{mainresult} is almost the same as Fig.\ref{Fig3}, which means the smoothed window function with smaller multipole values in harmonic space will bring smaller numerical error in E- and B-mode separation. We obtain the results: If $\delta_c $ is small, i.e. the less information loss, \textit{sin}- and \textit{cos}-smoothing methods are better than \textit{Gaussian}-smoothing method.
   On the other side, if we need the cleaner map, where $\delta_c$ should be larger (such as $\delta_c=1^\circ$ or $1.5^\circ$), \textit{Gaussian}-smoothing method is better. These can be understood by the following way: Comparing with the $sin$- or $cos$-smoothing functions, the $Gaussian$-smoothing function is much more steeper. So when $\delta_c$ is smaller, the $Gaussian$ function becomes close to the top-hat function, which will follow the larger numerical error. However, when $\delta_c$ is larger (i.e. $\delta_c>1^{\circ}$), all these three smoothing functions become relatively flat. While the $Gaussian$-smoothing function is continuous for any order derivatives around the boundaries, so the numerical errors can be deeply reduced in the numerical calculations. This can also explain why the leakage residuals in the constructed B-mode map are quite small when the $Gaussian$-smoothing function is adopted (see Fig.\ref{figurebfield}).


\section{conclusion}
\label{sect:conclusion}
\quad Detection of B-mode polarization is the main aim of the future CMB observations. For the real analysis, the incomplete sky survey induces the mixture of the E-mode and B-mode. In order to separate E- and B-mode of CMB on an incomplete sky, we need to smooth the edge of window function. In this article we present a comparison of the effects on numerical errors brought by different smoothing methods of the window function. We found that \textit{Gaussian}-smoothing method with large $\delta_c$ brings cleaner map, but also more information loss, while \textit{sin}- and \textit{cos}-smoothing methods do better when $\delta_c$ is small, i.e. less information loss.

\normalem
\begin{acknowledgements}
We acknowledge the use of the Planck Legacy Archive (PLA). Our data analysis made the use of HEALPix (\citealt{gorski2005healpix}) and GLESP (\citealt{doroshkevich2005gauss}). This work is supported by Project 973 under Grant No. 2012CB821804, by NSFC No. J1310021, 11173021, 11322324, 11421303 and project of KIP and CAS.

\end{acknowledgements}

\bibliographystyle{raa}
\bibliography{bibtex}

\begin{thebibliography}{26}
\providecommand\natexlab[1]{#1}
\providecommand\JournalTitle[1]{#1}

\bibitem[Ade {et~al.}(2015)]{ade2015joint}
Ade, P.~A., Aghanim, N., Ahmed, Z., {et~al.} 2015, Physical Review Letters,
  114, 101301

\bibitem[Ade {et~al.}(2014)]{ade2014evidence}
Ade, P., Akiba, Y., Anthony, A., {et~al.} 2014, Physical Review Letters, 112,
  131302

\bibitem[Aghanim {et~al.}(2015)]{aghanim2015planck}
Aghanim, N., Arnaud, M., Ashdown, M., {et~al.} 2015, arXiv preprint
  arXiv:1507.02704

\bibitem[Bock {et~al.}(2006)]{bock604101task}
Bock, J., {et~al.} 2006, arXiv preprint arXiv:astroph/0604101

\bibitem[Bunn(2008)]{bunn2008b}
Bunn, E.~F. 2008, arXiv preprint arXiv:0811.0111

\bibitem[Bunn {et~al.}(2003)]{PhysRevD.67.023501}
Bunn, E.~F., Zaldarriaga, M., Tegmark, M., \& de~Oliveira-Costa, A. 2003, Phys.
  Rev. D, 67, 023501

\bibitem[Cao \& Fang(2009)]{cao2009wavelet}
Cao, L., \& Fang, L.-Z. 2009, The Astrophysical Journal, 706, 1545

\bibitem[Doroshkevich {et~al.}(2005)]{doroshkevich2005gauss}
Doroshkevich, A., Naselsky, P., Verkhodanov, O.~V., {et~al.} 2005,
  International Journal of Modern Physics D, 14, 275

\bibitem[Efstathiou(2004)]{efstathiou2004myths}
Efstathiou, G. 2004, Monthly Notices of the Royal Astronomical Society, 349,
  603

\bibitem[Fert\'e {et~al.}(2013)]{PhysRevD.88.023524}
Fert\'e, A., Grain, J., Tristram, M., \& Stompor, R. 2013, Phys. Rev. D, 88,
  023524

\bibitem[Geller {et~al.}(2008)]{PhysRevD.78.123533}
Geller, D., Hansen, F.~K., Marinucci, D., Kerkyacharian, G., \& Picard, D.
  2008, Phys. Rev. D, 78, 123533

\bibitem[Gorski {et~al.}(2005)]{gorski2005healpix}
Gorski, K.~M., Hivon, E., Banday, A., {et~al.} 2005, The Astrophysical Journal,
  622, 759

\bibitem[Grain {et~al.}(2009)]{PhysRevD.79.123515}
Grain, J., Tristram, M., \& Stompor, R. 2009, Phys. Rev. D, 79, 123515

\bibitem[Hanson {et~al.}(2013)]{hanson2013detection}
Hanson, D., Hoover, S., Crites, A., {et~al.} 2013, Physical Review Letters,
  111, 141301

\bibitem[Kamionkowski {et~al.}(1997{\natexlab{a}})]{kamionkowski1997probe}
Kamionkowski, M., Kosowsky, A., \& Stebbins, A. 1997{\natexlab{a}}, Physical
  Review Letters, 78, 2058

\bibitem[Kamionkowski {et~al.}(1997{\natexlab{b}})]{PhysRevD.55.7368}
Kamionkowski, M., Kosowsky, A., \& Stebbins, A. 1997{\natexlab{b}}, Phys. Rev.
  D, 55, 7368

\bibitem[Kim(2011)]{kim2011make}
Kim, J. 2011, Astronomy \& Astrophysics, 531, A32

\bibitem[Kim \& Naselsky(2010)]{kim2010b}
Kim, J., \& Naselsky, P. 2010, Astronomy \& Astrophysics, 519, A104

\bibitem[Lewis(2003)]{PhysRevD.68.083509}
Lewis, A. 2003, Phys. Rev. D, 68, 083509

\bibitem[Lewis {et~al.}(2001)]{PhysRevD.65.023505}
Lewis, A., Challinor, A., \& Turok, N. 2001, Phys. Rev. D, 65, 023505

\bibitem[Naess {et~al.}(2014)]{naess2014atacama}
Naess, S., Hasselfield, M., McMahon, J., {et~al.} 2014, Journal of Cosmology
  and Astroparticle Physics, 2014, 007

\bibitem[Seljak \& Zaldarriaga(1997)]{seljak1997signature}
Seljak, U., \& Zaldarriaga, M. 1997, Physical Review Letters, 78, 2054

\bibitem[Smith(2006)]{PhysRevD.74.083002}
Smith, K.~M. 2006, Phys. Rev. D, 74, 083002

\bibitem[Smith \& Zaldarriaga(2007)]{PhysRevD.76.043001}
Smith, K.~M., \& Zaldarriaga, M. 2007, Phys. Rev. D, 76, 043001

\bibitem[Zaldarriaga \& Seljak(1997)]{PhysRevD.55.1830}
Zaldarriaga, M., \& Seljak, U. c.~v. 1997, Phys. Rev. D, 55, 1830

\bibitem[Zhao \& Baskaran(2010)]{PhysRevD.82.023001}
Zhao, W., \& Baskaran, D. 2010, Phys. Rev. D, 82, 023001

\end{thebibliography}
\end{document}